\documentclass{aa}
\usepackage{verbatim}
\usepackage{natbib}

\usepackage{amsmath}
\usepackage{amsfonts}
\usepackage{graphicx}
\usepackage{subcaption}
\usepackage{footnote}

\usepackage{txfonts}

\usepackage{hyperref}
\usepackage{cleveref}

\hypersetup{
colorlinks,
linkcolor={blue}, 
citecolor={blue},
urlcolor={blue},
}

\newcommand{\writetilde}{~}
\begin{document}

    \title{\texttt{Exoformer}: Accelerating Bayesian atmospheric retrievals with transformer neural networks}

    \author{L. Pagliaro\thanks{Email:leonardo.pagliaro@phd.unipd.it}\inst{\ref{inst1}}\,$^{\href{https://orcid.org/0009-0001-7886-949X}{\protect\includegraphics[height=0.19cm]{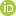}}}$
    \and T. Zingales\inst{\ref{inst1}, \ref{inst2}}\,$^{\href{https://orcid.org/0000-0001-6880-5356}{\protect\includegraphics[height=0.19cm]{Figures/orcid.jpg}}}$
    \and G. Piotto\inst{\ref{inst1}, \ref{inst2}}\,$^{\href{https://orcid.org/0000-0002-9937-6387}{\protect\includegraphics[height=0.19cm]{Figures/orcid.jpg}}}$ 
    \and I. Giovannini\inst{\ref{inst1},\ref{inst2}} \,$^{\href{https://orcid.org/0009-0005-9376-7044}{\protect\includegraphics[height=0.19cm]{Figures/orcid.jpg}}}$
    \and G. Mantovan\inst{\ref{inst3},\ref{inst2}}\,$^{\href{https://orcid.org/0000-0002-6871-6131}{\protect\includegraphics[height=0.19cm]{Figures/orcid.jpg}}}$}
    \authorrunning{L. Pagliaro et al.}

    \institute{Dipartimento di Fisica e Astronomia, Università degli Studi di Padova, Vicolo dell’Osservatorio 3, 35122 Padova, Italy\label{inst1}\and
    INAF, Osservatorio Astronomico di Padova, Vicolo dell’Osservatorio 5, 35122 Padova, Italy\label{inst2}\and
    Centro di Ateneo di Studi e Attivit\`a Spaziali ``G. Colombo'' -- Universit\`a degli Studi di Padova, Via Venezia 15, IT-35131, Padova, Italy\label{inst3}}

    \date{Received \dots, 2025; accepted March 09, 2026}
 
   \abstract
   {Computationally expensive and time-consuming Bayesian atmospheric retrievals pose a significant bottleneck for the rapid analysis of high-quality exoplanetary spectra from present and next generation space telescopes, such as JWST and Ariel. As these missions demand more complex atmospheric models to fully characterize the spectral features they uncover, they will benefit from data-driven analysis techniques such as machine and deep learning. We introduce and detail a novel approach that uses a transformer-based neural network (\texttt{Exoformer}) to rapidly generate informative prior distributions for atmospheric transmission spectra of hot Jupiters. We demonstrate the effectiveness of \texttt{Exoformer} using both simulated observations and real JWST data of WASP-39b and WASP-17b within the TauREx retrieval framework, leveraging the nested sampling algorithm. By replacing standard uniform priors with \texttt{Exoformer}-derived informative priors, our method accelerates nested-sampling retrievals by factor of 3-8 in the tested cases, while preserving the retrieved parameters and best-fit spectra. Crucially, we ensure that the retrieved parameters and the best-fit models remain consistent with results from classical methods. Furthermore, we confirm the statistical consistency of the two retrieval approaches by comparing their log-Bayesian evidence, obtaining absolute values of each Bayes factor $|\Delta\log{Z}|<5$, i.e., with no strong preference following common scales for either model. This hybrid approach significantly enhances the efficiency of atmospheric retrieval tools without compromising their accuracy, paving the way for more rapid analysis of complex exoplanetary spectra and enabling the integration of more realistic atmospheric models.}

    
    \maketitle

    \section{Introduction}

    The advent of the James Webb Space Telescope (JWST) \citep{Gardner_2006} and forthcoming Ariel \citep{Tinetti_2021} space missions opens the door to unprecedented spectroscopic observations of exoplanetary atmospheres. The study of atmospheric compositions plays a crucial role in understanding how planets form and evolve, as well as in detecting molecular signatures of life. Many tools based on Bayesian statistics have been developed to retrieve molecular abundances, temperatures, and many other atmospheric parameters from spectroscopic observations (e.g., \texttt{TauREx} \citep{AlRefaie2021}, \texttt{NEMESIS} \citep{Irwin_2008}, and \texttt{petitRADTRANS} \citep{Molliere_2019}; a more complete list can be found in \citet{MacDonald_Batalha_2023}). The classical approach involves a Bayesian framework to obtain the posterior distributions of atmospheric parameters given an observed spectrum. These tools prove to be effective with low-resolution observations of the HST and Spitzer telescopes. However, high-quality data from JWST and Ariel show a completely new set of spectral features that can be described only with more complex atmospheric models \citep{Rocchetto_2016}, considering for example atmospheres as multidimensional structures where physical phenomena such as convection and chemical disequilibrium occur. The resulting increase in parameters describing the atmospheric model yields a strong computational bottleneck and challenges in achieving convergence to a solution.\par
    In recent years, an increasing number of machine learning and deep learning algorithms have been implemented in the exoplanetary field, ranging from transit detection in light curves (e.g., \citet{McCauliff2015}, \citet{Shallue2018}) to atmospheric characterization (e.g., \citet{Yip_2021}, \citet{Himes2022}, \citet{Vasist_2023}). These algorithms are based on a data-driven approach, which means that they can automatically learn to solve a given task by iteratively extracting multiple features from a large dataset \citep{Janiesch2021}. Machine learning encompasses a broad range of algorithms derived from different learning paradigms (e.g., decision trees \citep{Quinlan_1986}, support vector machines \citep{Cortes_Vapnik_1995}, and clustering \citep{Kaufman_2005}). Deep learning by contrast relies on the artificial neuron -- a mathematical function that employs nonlinear transformations through weighted inputs to process and learn from data -- and its variants (e.g., convolution and attention). Individual neurons are typically grouped into distinct layers that are then repeated and connected to form more complex architectures, called deep neural networks. Deep neural networks are used extensively in complex tasks (such as images, texts, and sequential data processing), as they outperform machine learning in extracting hidden features and patterns from domains with large, high-dimensional data \citep{LeCun_2015, Janiesch2021}.\par
    Bayesian retrievals require millions of atmospheric forward models per single observation, becoming extremely slow and computationally intensive when complex atmospheric models with a large number of parameters are taken into account. Therefore, with the thousands of spectra expected from new telescopes, the efficiency of repeated analysis will be significantly impacted. The integration of deep learning emerges as a promising solution to address the computational challenges of high-dimensional Bayesian atmospheric retrievals. There are multiple pathways through which we can achieve this goal: replacing the Bayesian tool (e.g., \citet{Zingales2018}) to directly provide parameter posterior distributions, substituting the radiative transfer code with a surrogate model (e.g., \citet{Himes2022}), or generating informative priors (e.g., \citet{Hayes2020}) in place of uniform priors.\par
    In our work, we adopted a state-of-the-art deep-learning architecture, the transformer, whose architecture was proposed by \citet{vaswani2017attention} in the context of human language modeling to overcome the limitations of convolutional neural networks (CNN; \citep{LeCun_1989}) and long-short term memory (LSTM; \citet{Hochreiter_Schmidhuber_1997}) networks. Building on the success of transformers in predicting stellar parameters – as demonstrated in recent studies (e.g., \citet{Pan2024}, \citet{Zhang2024}) – we propose a transformer-based approach for analyzing spectroscopic data of exoplanetary atmospheres. Just as stellar spectra exhibit interconnected emission and absorption features across the spectral domain, a transformer architecture is well suited to capture these complex relationships within exoplanetary atmospheric data. We developed a tool that can retrieve the approximated posterior distributions of six atmospheric parameters using the Monte Carlo (MC) dropout technique \citep{Gal&Ghahramani2016}. In Section \ref{section:transformer} we describe the transformer architecture and its fundamental operation, self-attention. In Section \ref{section:exoformer} we introduce \texttt{Exoformer}, our transformer-based tool, and describe its structure. In Section \ref{section:methods} we describe the training process of \texttt{Exoformer} and how uncertainties are estimated in the neural network. In Sections \ref{section:results} and \ref{section:informative-priors}, we show the results of applying \texttt{Exoformer} to simulated and real JWST transmission spectra.

    \section{Transformer neural networks}
    \label{section:transformer}
    
    Transformers are a type of neural network designed to learn useful representations of sequential data through a mechanism called self-attention. In fact, unlike other architectures such as CNN and LSTM, transformers have the ability to capture long-range dependencies and correlations in sequences. In the following paragraphs, we describe the most important components of a transformer algorithm.

    \subsection{Self-attention mechanism}

    Self-attention is the core mechanism of the transformer architecture: it allows the model to relate each element of a sequence to all the other elements of the same sequence. In general, the $i$-th output $\mathbf{a}_i\in\mathbb{R}^D$ (where $D$ is the embedding dimension) of the self-attention operation is a weighted sum of the $N$ inputs $\mathbf{x}_1, \dots, \mathbf{x}_N$, with $\mathbf{x}_j\in\mathbb{R}^D$:
    \begin{equation}
        \mathbf{a}_i = \sum_{j=1}^{N} {A}_{ij} \mathbf{x}_j,
        \label{eq:self-attention}
    \end{equation}
    where $\mathbf{A}\in\mathbb{R}^{N\times N}$ is the attention matrix, whose elements are normalized between $[0, 1]$ and their sum equals to $1$. \\
    \citet{vaswani2017attention} applied the self-attention mechanism to deep learning by introducing a set of learnable matrices to compute the attention matrix and the attention output. Given the sequence of $N$ inputs $\mathbf{x}_1, \dots , \mathbf{x}_N\in\mathbb{R}^D$, each vector is first linearly transformed by three distinct learnable matrices $\mathbf{W}\in\mathbb{R}^{D\times D}$:
    \begin{align}
        \mathbf{q}_i = \mathbf{x}_i\mathbf{W}_Q,\\
        \mathbf{k}_i = \mathbf{x}_i\mathbf{W}_K,\\
        \mathbf{v}_i = \mathbf{x}_i\mathbf{W}_V.
    \end{align}
    The three resulting vectors $\mathbf{q}_i, \mathbf{k}_i, \mathbf{v}_i\in\mathbb{R}^{D}$ (called query, key, and value, respectively) are then used to obtain the $\mathbf{A}_{ij}$ element of the attention matrix through
    \begin{equation}
         {A}_{ij} = \text{softmax}(\mathbf{q}_i\mathbf{k}^T_{j}) = \frac{\exp{\big(\mathbf{q}_i\mathbf{k}^T_{j}\big)}}{\sum_{n=1}^{N}\exp{\big(\mathbf{q}_{i}\mathbf{k}^T_{n}\big)}}
         \label{eq:attention-matrix}
    \end{equation}
    and the self-attention output given by
    \begin{equation}
        \mathbf{a}_i = \sum_{j=1}^{N}{A}_{ij}\mathbf{v}_j.
    \label{eq:self-attention-nn}
    \end{equation}
    In Eq.\eqref{eq:attention-matrix} every query $\mathbf{q}_i$ is compared to all $N$ keys $\mathbf{k}_j$ to find the combinations with the highest correlation by using the geometric properties of the dot product. Equation \eqref{eq:self-attention-nn} instead creates a new representation of the value $\mathbf{v}_i$, where the information about the pairs of vectors with the highest attention scores is stored.\\
    Stacking the $N$ input vectors into a column matrix $\mathbf{X}\in\mathbb{R}^{N\times D}$, we can rewrite Eq.\eqref{eq:self-attention-nn} in a simpler way:
    \begin{equation}
    \mathbf{a} = \text{softmax}(\mathbf{Q}\mathbf{K}^T)\mathbf{V},
    \end{equation}
    where $\mathbf{Q}\in\mathbb{R}^{N\times D}, \mathbf{K}\in\mathbb{R}^{N\times D}, \mathbf{V}\in\mathbb{R}^{N\times D}$ are the query, key, and value matrices for the entire sequence.\\
    Moreover, \citet{vaswani2017attention} improved the self-attention numerical stability by scaling the argument of the softmax with the square root of the embedding dimension:
    \begin{equation}
        \mathbf{a} = \text{softmax}\bigg(\frac{\mathbf{Q}\mathbf{K}^T}{\sqrt{D}}\bigg)\mathbf{V}.
    \end{equation}

    \nocite{prince2023understanding}
    \nocite{tanoglidis2023transformers}
    \nocite{Turner2024}

    \subsection{Multi-head self-attention}
    \label{sec:mhsa}

    Multiple self-attention mechanisms are typically applied in parallel to capture further relationships between the input vectors of a sequence. This approach introduced by \citet{vaswani2017attention} is called multi-head self-attention.\\
    Consider $H$ self-attention heads $\mathbf{a}_h$, each with a different set of queries $\mathbf{Q}_h$, keys $\mathbf{K}_h$, and values $\mathbf{V}_h$: every head acts on a fraction ${{D}/{H}}$ of the input embedding, allowing the extraction of correlations from different parts of the embedding. The output of all self-attention heads are then concatenated, and a final linear transformation $\mathbf{W}_o\in\mathbb{R}^{D\times D}$ is applied:
    \begin{equation}
        \mathbf{mhsa} = \text{Concat}(\mathbf{a}_1, \dots, \mathbf{a}_h)\mathbf{W}_o.
    \end{equation}
    This multiheaded attention mechanism is introduced to further enhance the performance of single self-attention by capturing deeper and more specific correlations in input data.
    
    \subsection{Positional encoding}
    \label{sec:positional-encoding}
    
    By definition, the self-attention mechanism is invariant under permutation of the input sequence $\mathbf{X}$: by permuting the columns of $\mathbf{X}$, we permute all its representations across the mechanism in the same way. However, positional information is a key aspect in spectroscopy, because it is related to the different absorption or emission features of molecules. \par
    To fix the problem, we can add a vector $\mathbf{p}_i\in\mathbb{R}^D$ to the column vector $\mathbf{x}_i$,
    \begin{equation}
        \mathbf{x}_i \rightarrow \mathbf{x}_i + \mathbf{p}_i,
    \end{equation}
    where $\mathbf{p}_i$ can be a customized mathematical function or a parameter learned during training. The vector $\mathbf{p}_i$ encodes the position information inside the embedding, before self-attention is applied: for this reason, we refer to this mechanism as positional encoding.\par
    \citet{vaswani2017attention} presented a positional encoding based on a combination of sine and cosine functions:
    \begin{align}
        \mathbf{p}_i(2j+1) = \cos{\left(\frac{i}{10000^{2j/D}}\right)}, \\
        \mathbf{p}_i(2j) = \sin{\left(\frac{i}{10000^{2j/D}}\right)},
    \end{align}
    where $j$ indicates the elements of the vector. This representation ensures a unique position is associated with each element of the input sequence, while maintaining the output in a fixed range.
    
    \subsection{Encoder block}
    
\begin{figure}[ht!]
      \centering
      \includegraphics[trim={9.1cm 0.0cm 5.5cm 0.0cm},clip,width=10cm]{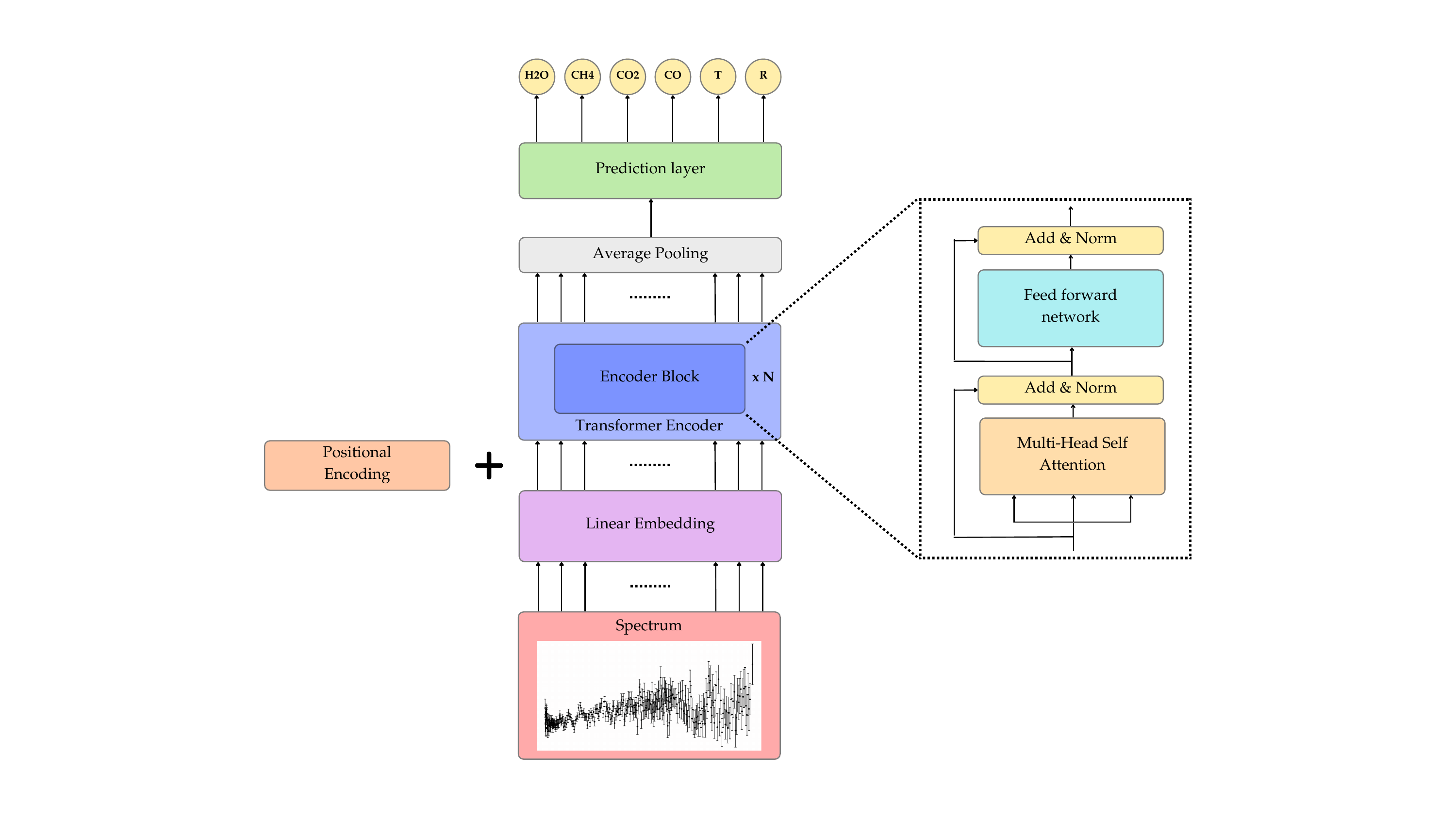} 
      \caption{Schematic of the \texttt{Exoformer} architecture. Each box represents a layer described in Section \ref{section:exoformer}. Inside the dashed box, the layers forming a single encoder block are indicated. Multiple encoder blocks repeated sequentially form the transformer encoder.}
      \label{fig:transformer-layer}
    \end{figure}

    Sections \ref{sec:mhsa} and \ref{sec:positional-encoding} introduced the two main elements of the transformer architecture. In our work, we used only the encoder part of the architecture described in \citet{vaswani2017attention}. The transformer encoder was built with a series of encoder blocks, each containing the following layers (dashed box in Figure \ref{fig:transformer-layer}):
    \begin{enumerate}
        \item  multi-head self-attention;
        \vspace{10pt}
        \item  a residual (skip) connection around the $\mathbf{mhsa}$ output, where the output $\mathbf{Y}$ is written as
        \begin{equation}
            \mathbf{Y}=\mathbf{X}+\mathbf{mhsa}(\mathbf{X});
        \end{equation}
        \item  layer normalization that standardizes the skip-connection output to zero mean and unit variance, improving the transformer's numerical stability;
        \vspace{10pt}
        \item  a feed-forward neural network applied to each output vector of layer normalization;
        \vspace{10pt}
        \item an additional skip-connection and final layer normalization.
    \end{enumerate}
    In general, the encoder block is repeated multiple times inside a transformer encoder, receiving as input the output of the previous encoder block. This sequence of encoders enables the extraction of further correlations within the data.

    \section{Exoformer}
    \label{section:exoformer}
    
    We now present our transformer based tool, \texttt{Exoformer}, used to infer the values of six atmospheric parameters based on the data provided by the publicly available training dataset\footnote{\href{https://osf.io/6dxps/files/osfstorage}{https://osf.io/6dxps/files/osfstorage}} by \citet{Zingales2018}. The transformer encoder forms the core of our model. However, other types of layers were integrated to extract the six predicted atmospheric parameters from an input transmission spectrum. Fig.~\ref{fig:transformer-layer} illustrates \texttt{Exoformer}'s structure: the shaded boxes indicate different layers, each with a specific functionality.

    \subsection{Linear embedding}

    An input transmission spectrum with $515$ spectral points first passes through a linear embedding layer. There, each spectral point is transformed into a $128$D vector using a multilayer perceptron (MLP), which yields a ${515\times128}$ $2$D output. The first dimension represents the wavelength grid, while the new axis encodes preliminary spectral features captured by the MLP during the learning process. Lastly, the positional encoding vector is added to the embedding vector.
    
    \subsection{Transformer encoder}

    The resulting vector is then injected into the transformer encoder. This component is characterized by multiple transformer layers arranged in series and captures the long-range relations between the embeddings. \texttt{Exoformer} uses five transformer encoder blocks, each comprising eight multi-head self-attention layers and a feed-forward network with $1024$ hidden units. Since the output of each transformer layer has the same dimensions as the input, the resulting vector from the transformer encoder has a $515\times 128$ shape.
    
    \subsection{Prediction layer}
    
    The last layer of \texttt{Exoformer} is used to predict the values of the six atmospheric parameters. To do this, we first applied four average pooling layers along the embedding dimension. The resulting vector provides an average representation of each embedding and is finally passed to a two-layer MLP, which produces six scalar outputs.
    
    \section{Methods}
    \label{section:methods}

    In this section, we discuss in detail the training procedure for \texttt{Exoformer}. We adopted the standard methodology for developing a deep learning tool, which includes using a representative dataset of real-world scenarios, applying data normalization for numerical stability, finding hyperparameters for the best model, and performing uncertainty estimation for comparison with Bayesian tools.
    
    \subsection{Dataset}
    \label{sec:dataset}
    
    To train \texttt{Exoformer}, we relied on the dataset described in \citet{Zingales2018}. The dataset contains $10^7$ atmospheric transmission spectra of hot Jupiters, generated using the analytical forward model of TauREx 3 \citep{AlRefaie2021}. Each spectrum was parameterized with seven atmospheric parameters: four molecule abundances (H$_2$O, CH$_4$, CO, and CO$_2$), isothermal temperature $T_{iso}$, and the planet's mass $M_p$ and radius $R_p$. The atmospheric forward model, in addition to the absorption contributions of the four molecules, includes Rayleigh scattering and collisionally induced absorptions (CIAs) from H$_2$-H$_2$ and H$_2$-He pairs. Table \ref{table:1} summarizes the upper and lower boundaries of the seven parameters. All spectra were binned to a custom wavelength grid, ranging from $0.3$ to $50$ $\mu m$, with $515$ spectral points. This allowed us to cover not only the main absorption features of the four previously mentioned molecules but also the principal band passes of JWST and Ariel. For our regression problem, we used six of these seven available parameters: H$_2$O, CH$_4$, CO, and CO$_2$ abundances, the isothermal temperature $T_{iso}$, and the planet's radius $R_p$.\par
    We then divided the entire dataset into three distinct subsets: training, validation, and test datasets. The training set was used to train the network, the validation set monitored the learning progress, and the test set identified the model with the best performance. The proportion assigned to each set is as follows: $90\%$ for training, $9\%$ for testing, and $1\%$ for validation. Following modern best practices for large-scale datasets ($10^7$ spectra), a $1\%$ validation split provided a statistically significant sample size of $10^5$ instances. This proportion ensured a reliable estimate of model performance while optimizing computational efficiency during the training process.
    
    \begin{table}[h!]
    \def\arraystretch{1.5}
    \centering
    \caption{Boundary values of the seven atmospheric parameters used to generate the training dataset in \citet{Zingales2018}.}
    \label{table:1}
    \begin{tabular}{lcc}
    \hline\hline
    Parameter & Lower Bound & Upper Bound \\
    \hline
    $\log$ H$_2$O & $-8$ & $-1$ \\
    $\log$ C$H_4$ & $-8$ & $-1$ \\
    $\log$ CO & $-8$ & $-1$ \\
    $\log$ C$O_2$ & $-8$ & $-1$ \\
    $M_p$ & $0.8$ $M_J$ & $2.0$ $M_J$\\
    $R_p$ & $0.8$ $R_J$ & $1.5$ $R_J$\\
    $T_{iso}$ & $1000$ $K$ & $2000$ $K$\\
    \hline
    \end{tabular}
    \end{table}

    \begin{figure*}[t]
      \centering
      \sidecaption
      \includegraphics[trim={0.1cm 0.3cm 0.1cm 0.1cm},clip,width=12cm]{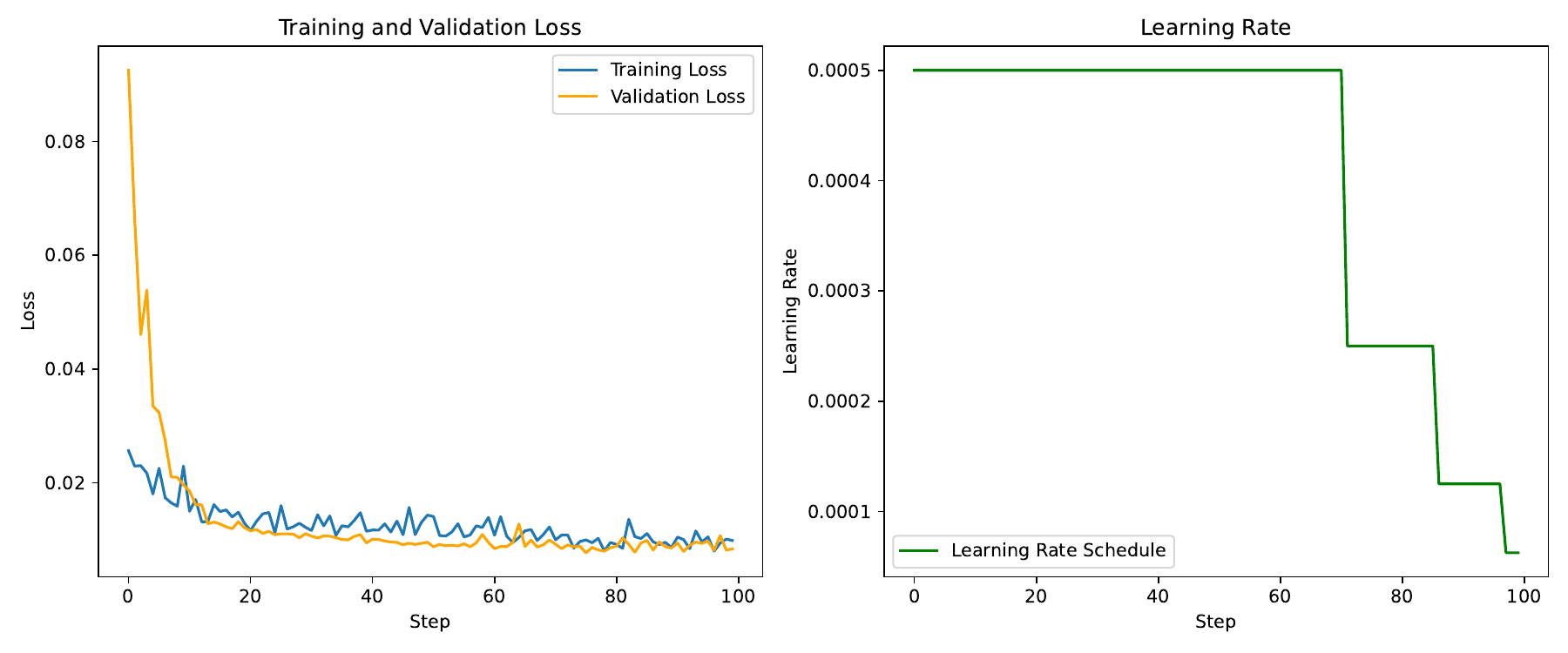}
      \caption{Left plot: Training and validation losses as a function of the training step. Right plot: Learning rate trend as determined by the learning rate schedule applied during training.}
      \label{fig:losses}
    \end{figure*}
    
    \subsection{Spectra preprocessing} \label{subsubsec:pre-processing}
    
\begin{figure}[t]
    \sidecaption
      \centering
      \resizebox{\hsize}{!}{\includegraphics{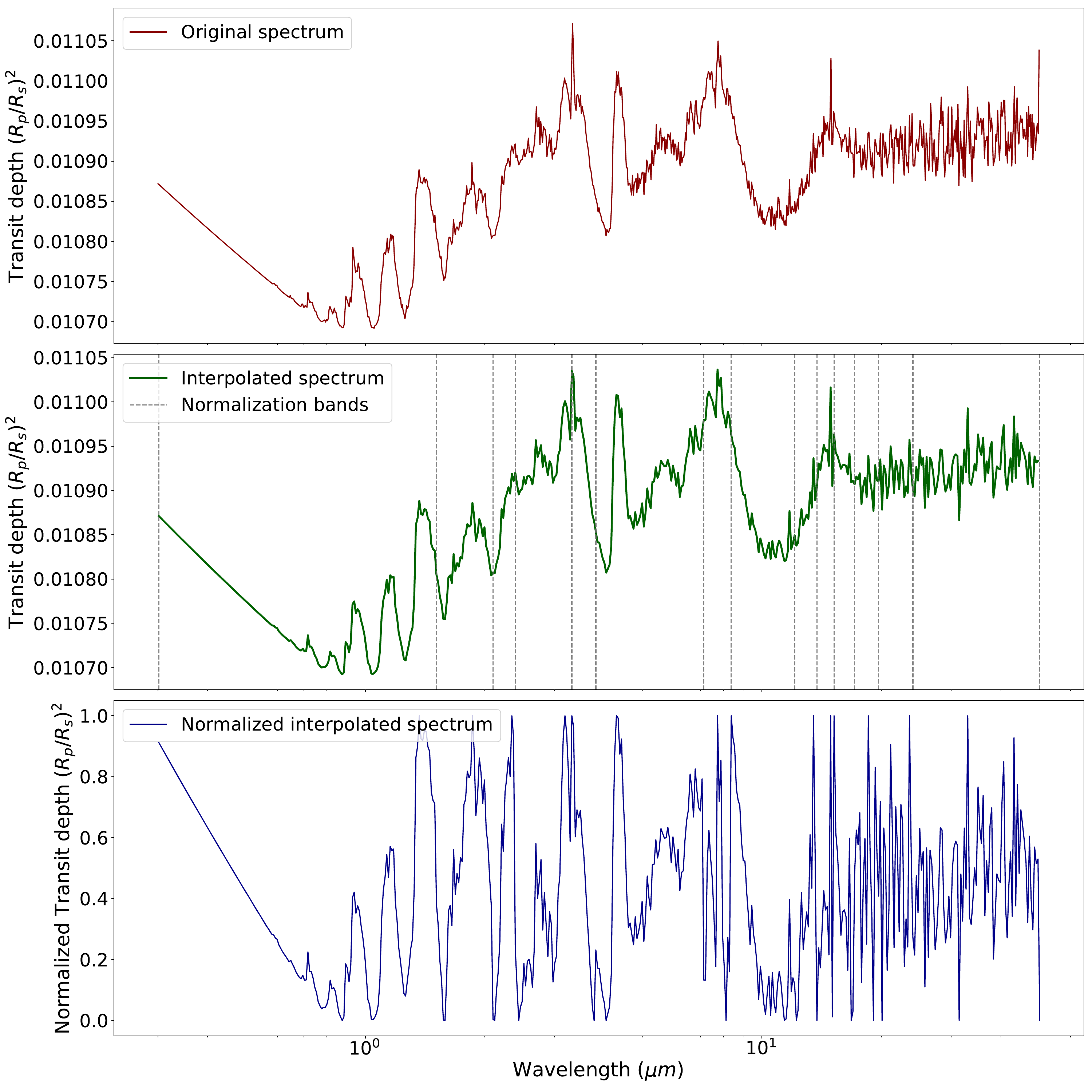}}
      \caption{Preprocessing phases on the test planet spectrum in Table \ref{table:test-case-planet}. Upper plot: Analytical spectrum computed using the TauREx forward model. Middle plot: Analytical spectrum binned to the custom grid and normalization bands. Bottom plot: Interpolated spectrum after normalization.}
      \label{fig:preprocessing}
    \end{figure}

    Because transmission spectra exhibit varying transit depth magnitudes, we need to scale each spectrum to a consistent range. This prevents \texttt{Exoformer} from assigning higher importance to inputs with higher raw values -- a situation we want to avoid to ensure the tool remains unbiased. Consequently, we applied the normalization scheme introduced in \citet{Zingales2018}: every spectrum was divided into 14 bands, and the spectral points in each interval were normalized between $0$ and $1$ using the maximum and minimum values of those intervals. Figure~\ref{fig:preprocessing} illustrates how a spectrum is transformed after applying the normalization scheme. The second panel depicts the 14 wavelength bands as vertical dashed lines, while the third shows that, within each subinterval defined by these bands, the spectral point with the highest transit depth normalizes to 1 and the point with the lowest transit depth normalizes to 0.\par
    To mitigate similar biases in the outputs, atmospheric parameters values were also normalized between $0$ and $1$. We applied min-max scaling using the bounds in Table\writetilde\ref{table:datasetboundaries} for each parameter x:
    \begin{equation}
        z=\frac{(x-\text{min})}{(\text{max}-\text{min})},
    \end{equation}
    where $z$ is the scaled $x$ parameter, while min and max are the lower and upper boundaries, respectively.

    \subsection{Training}

    \texttt{Exoformer} was trained on the training dataset using the AdamW optimizer \citep{loshchilov2017decoupled} with mini-batches of size $64$, initial learning rate of $0.5\cdot10^{-4}$, and weight decay of $1\cdot10^{-4}$. To prevent \texttt{Exoformer} training from stagnating in the loss landscape, we used a scheduler that reduces the learning rate by a factor of $0.5$ if the validation loss does not improve over ten consecutive steps (Fig.\writetilde\ref{fig:losses}, right panel). Since \texttt{Exoformer} was used for a regression problem, we chose the mean squared error (MSE) between the real and predicted atmospheric parameters as training loss. In the left plot of Fig.\writetilde\ref{fig:losses} training and validation losses are shown  as functions of training steps, illustrating model convergence and generalization capability. The best model hyperparameters were found using the Optuna \citep{optuna2019} Python library, with root mean squared error (RMSE) as the performance metric to minimize. 
    
    \subsection{Uncertainty estimation}
    
    Traditional deep-learning regression approaches focus on generating a single predicted value for each output parameter. To obtain a measure of the output uncertainty, which is crucial for our work, techniques such as MC dropout \citep{Gal&Ghahramani2016} are used. These techniques allow multiple outputs to be sampled during the inference phase, effectively providing a distribution of possible predictions. Our choice to use MC dropout is motivated by its computational efficiency and ease of implementation. In \texttt{Exoformer} the MC dropout mechanism was applied to all layers containing dropout.
    
    Monte Carlo dropout arises from the need to predict the posterior distribution of an output $y^*$ given some unseen data $x^*$ with a neural network $f$ defined by a set of weights $\mathbf{W}$,
    \begin{equation}
        \mathbf{y}^*=f(\mathbf{x}^*, \mathbf{W}),
    \end{equation}
    trained on a dataset $D=\{\mathbf{X}, \mathbf{Y}\}=\{x_1, \dots, x_N, y_1, \dots, x_N\}$. The expression of the posterior distribution is given by Bayesian inference:
    \begin{equation}
        P(y^*|x^*,D) = \int P(y^*|x^*, \mathbf{W})P(\mathbf{W}|D)\, dW.
        \label{eq:predictive-distribution}
    \end{equation}
    \citet{Gal&Ghahramani2016} found that the previously defined neural network is equivalent to an approximation of the Gaussian process when dropout regularization is applied. This allows rewriting Eq.\eqref{eq:predictive-distribution} as
    \begin{equation}
        P(y^*|x^*,D)\approx q(y^*|x^*)\approx \mathcal{N}(f(x^*, W)).
    \end{equation}
    In this way, we can compute the predictive mean and variance by running $N_{step}$ forward passes of the neural network with dropout regularization kept active during inference time (after the training phase, when the neural network is used for predictions). The resulting expressions for mean and variance are, respectively,
    \begin{equation}
        \text{Mean}_{q(y^*|x^*)}(y^*)=\mathbb{E}_{q(y^*|x^*)}(y^*)\approx\frac{1}{N_{step}}\sum_{i=1}^{N_{step}} f(x^*, \mathbf{W}),
    \end{equation}
    \begin{equation}
        \begin{split}
            \text{Var}_{q(y^*|x^*)}(y^*) \approx &\, \frac{1}{N_{step}}\sum_{i=1}^{N_{step}}f(x^*, \mathbf{W})^Tf(x^*, \mathbf{W})\\
            &-\mathbb{E}_{q(y^*|x^*)}(y^*)^T\mathbb{E}_{q(y^*|x^*)}(y^*).
        \end{split}
        \label{eq:variance}
    \end{equation}
    In other words, the variance includes not only the uncertainty of the model but also the uncertainty from the training data.

    \section{Results}
    \label{section:results}

    In this section, we analyze the performance of \texttt{Exoformer} trained on the \citet{Zingales2018} dataset. We evaluated its effectiveness by assessing its performance on a simulated JWST observation of a hot Jupiter and confirmed its robustness using unseen data. Subsequently, we compared its results with those from a Bayesian retrieval tool applied to real JWST observational data.
    
    \subsection{Retrieval on simulated observations}
    \label{subsec:testretrieval}
    
\begin{figure*}[t]
    \sidecaption
      \centering
      \includegraphics[trim={1cm 0.5cm 3.5cm 2.2cm},clip,width=12cm]{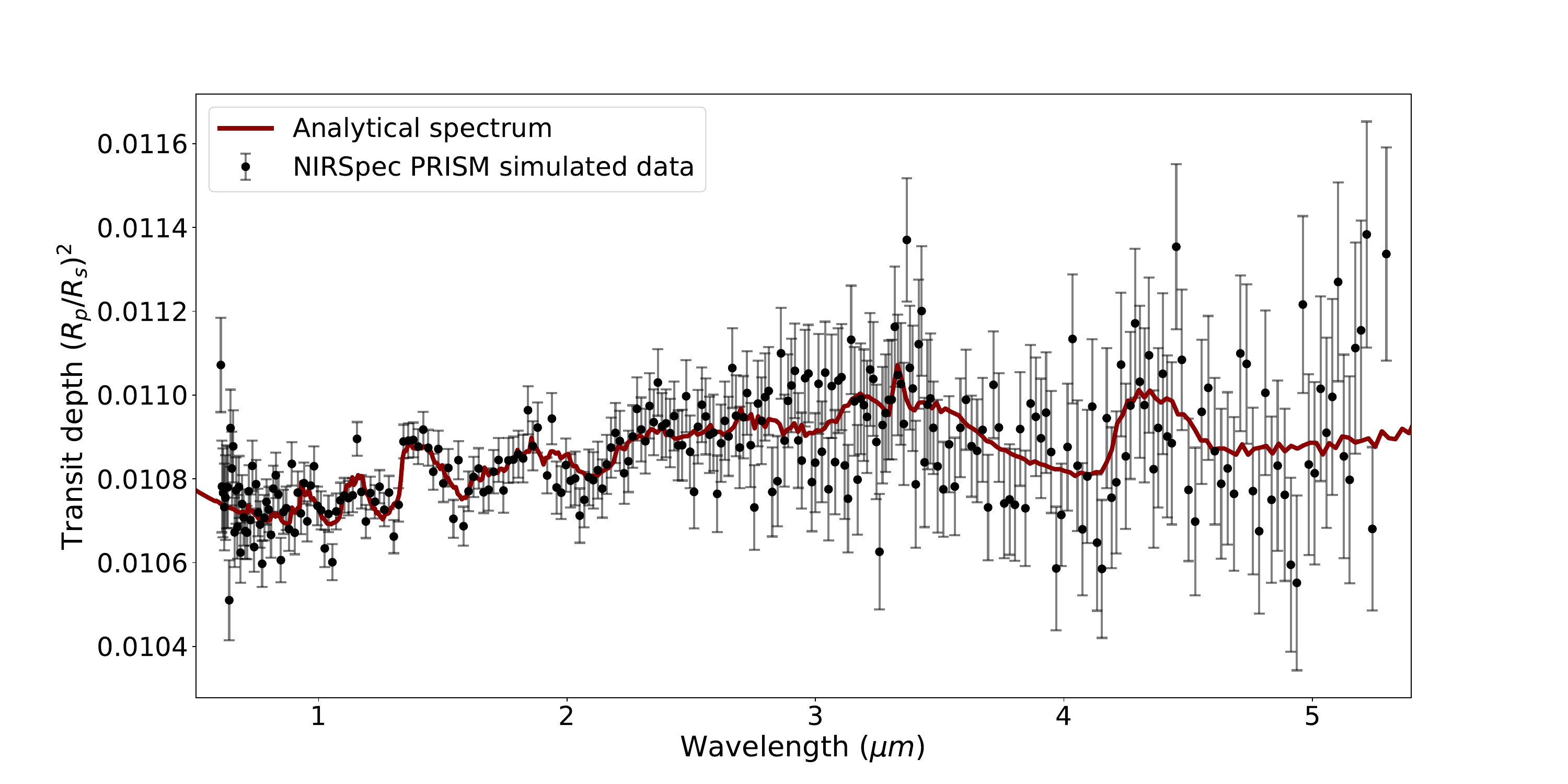}
      \caption{Simulated NIRSpec PRISM observation of the transmission spectrum in Fig.~\ref{fig:preprocessing}. The observational data points (black dots) are binned to the native resolution of NIRSpec PRISM ($R=100$) and superimposed on the original (red line) TauREx analytical spectrum.}
      \label{fig:nirspec-simulation}
    \end{figure*}

    To evaluate the retrieval performance of \texttt{Exoformer} using realistic data, we used the atmospheric model from TauREx 3 \citep{AlRefaie2021} to simulate the transmission spectrum of a hot Jupiter. The model includes the absorption contributions from the four chemical species mentioned above (with cross sections from \texttt{ExoMol} \citep{Tennyson_2024}), an isothermal temperature profile, Rayleigh scattering, and CIA contributions. A summary of the reference values for the planet is provided in Table \ref{table:test-case-planet}.
    We then used \texttt{Pandexo} \citep{Batalha2017}, a tool for simulating JWST spectroscopic observations of exoplanetary atmospheres to create a more realistic observation of the test exoplanet. The simulated observation includes a noise floor of 30 ppm, a transit duration of four hours, and a single transit of the planet. For this simulation, we selected the NIRSpec instrument operating in PRISM mode, which covers a wavelength range of $0.7-5.0$ $\mu m$ at a native resolution of 100. In Fig.~\ref{fig:nirspec-simulation} the analytical spectrum generated with TauREx (red line) and the NIRSpec Prism observation simulation (dots with error bars) are shown.

    To account for uncertainties in real observations, we assumed Gaussian-distributed errors. Doing so, we generated a set of noisy spectra $\mathbf{x}_i(\lambda)$ by sampling $N_{\text{sample}}$ times from a normal distribution. The mean value of the distribution for each spectral point corresponds to the observed transit depth at wavelength $\lambda_j$, while the standard deviation is equal to the associated error at $\lambda_j$. We interpolated each noisy spectrum to the \texttt{Exoformer} grid using a cubic scheme, setting all points outside the instrument coverage to zero. Thanks to the correlations between the spectral features (captured during training and stored in the embedding), the transformer can still make predictions even when spectral data from wavelengths outside the instrument bands is missing. As a last step, we applied the normalization scheme described in \ref{subsubsec:pre-processing}.

    \begin{figure*}[h!]
    \sidecaption
      \centering
      \includegraphics[trim={0.2cm 0.2cm 0.5cm 0.5cm},clip,width=12cm]{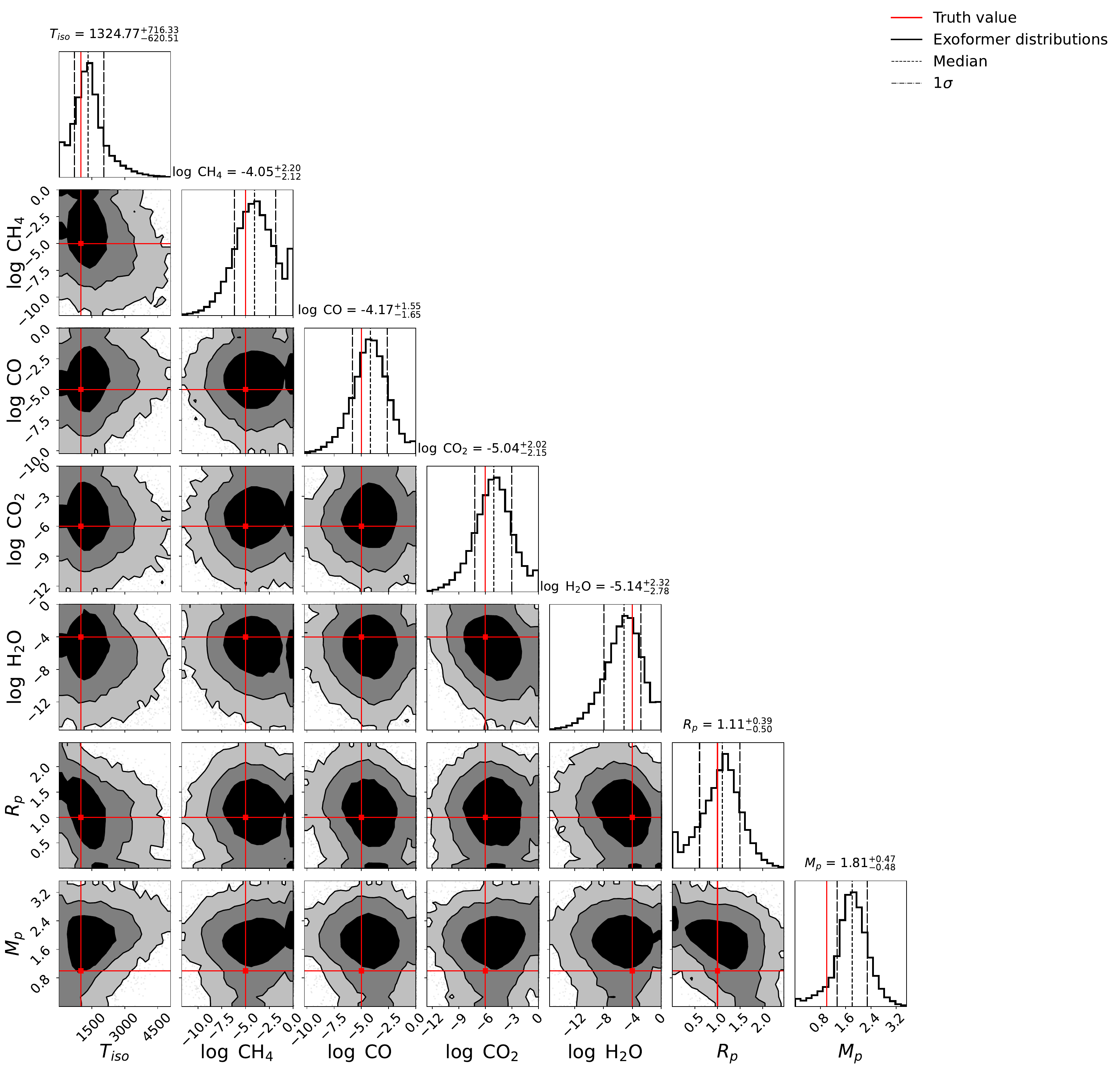}
      \caption{Posterior distributions and ground truth values (red lines) of the seven parameters. The retrieval was performed using \texttt{Exoformer} on the NIRSpec PRISM simulation. The dashed lines indicate the median of the distribution, while the dashed-dotted lines indicate the $1\sigma$ intervals.}
      \label{fig:original-vs-nirspec}
    \end{figure*}
    
    We performed inference on $N_{\text{sample}}=500$ noisy spectra samples, applying MC dropout with $N_{\text{step}}=100$ and $p_{\text{drop}}=0.8$ to each. The $N_{\text{step}}$ value was chosen to avoid excessive computational overhead during inference, while $p_{\text{drop}}$ was treated as a hyperparameter and optimized to keep \texttt{Exoformer} consistent with the ground truth values of the atmospheric parameters. In total, we obtained a set of $N_{\text{sample}}$ predictive distributions, each with $N_{\text{step}}$ elements, for the seven parameters. Finally, these distributions were concatenated to compute the mean and $1\sigma$ bounds of the parameters, as described in \citet{Gal&Ghahramani2016}. Figure~\ref{fig:original-vs-nirspec} shows the retrieval results: \texttt{Exoformer} predictions are consistent (except for mass) with the ground truth values in Table \ref{table:test-case-planet} within the error bars. The distributions are Gaussian-like, as expected from the definition of MC dropout. The width of the distributions is controlled by the variance in Eq.\eqref{eq:variance}, which is proportional to the dropout probability \citep{Gal&Ghahramani2016}. Since we used a high value for this probability, we expect a high value for the $1\sigma$ intervals. Another aspect is the increase of the distribution toward the edge of the boundary conditions, which arises from \texttt{Exoformer}'s prediction layer. In fact, a ReLU activation function was applied to the last layer. Since its expression is $max(0, x)$, it sets all negative values to zero without limiting positive values from the previous layer.\par
    Planetary mass retrieval is challenging due to its degeneracy with other atmospheric parameters \citep{Changeat_2020}, as distinct parameter combinations can produce similar spectra. Consequently, to mitigate the impact of the planetary mass uncertainty on the accuracy of atmospheric composition \citep{DiMaio_2023}, we did not include the mass prediction from \texttt{Exoformer} and instead fixed its value in subsequent retrievals.
    
    \begin{table}
    \centering
    \caption{Planetary parameters for the test case planet used as input for the TauREx forward model.}
    \label{table:datasetboundaries}
    \def\arraystretch{1.5}
    \begin{tabular}{lc}
    \hline\hline
    \multicolumn{2}{c}{\textbf{Planet}} \\
    \hline
    $\log{\text{H$_2$O}}$ & $-4$ \\
    $\log{\text{CH$_4$}}$ & $-5$ \\
    $\log{\text{CO}}$ & $-5$ \\
    $\log{\text{CO$_2$}}$ & $-6$ \\
    $M_p$ & $1$ $M_{J}$\\
    $R_p$ & $1$ $R_{J}$ \\
    $T_{iso}$ & $1000$ $K$\\
    \hline
    \end{tabular}
    \label{table:test-case-planet}
    \tablefoot{The host star is a Sun-like, as used by \citet{Zingales2018} in the training dataset.}
    \end{table} 

    \subsection{Comparison with Bayesian retrieval tools}
    \label{sec:comparisontaurex-exoformer}

    We then tested \texttt{Exoformer} against a Bayesian retrieval tool on real JWST transmission spectra. For this comparison, we selected WASP-39b and WASP-17b, two hot Jupiter with parameters falling within the limits of the training dataset. We used the transmission spectrum obtained with the NIRSpec instrument in PRISM mode, reduced with the FIREFly pipeline by \citet{Rustamkulov2023} for WASP-39b, and the transmission spectrum from NIRISS in SOSS mode for WASP-17b reduced by \citet{Louie2025}.
    
     We first performed the retrieval with TauREx, applying the same forward model used to generate the training dataset and the analytical spectrum of the test planet. The model includes six fitting parameters, corresponding to those retrieved by \texttt{Exoformer}. We assigned log-uniform prior ranges of $10^{-10}-10^{-1}$ to the molecular mixing ratios, whereas we applied uniform distributions for the priors of $T_{iso}$ and $R_p$, with ranges $1000-2000$ $K$ and $0.8-1.5$ $R_J$, respectively. Following the procedure described in Section \ref{subsec:testretrieval}, we performed the analysis of WASP-39b and WASP-17b with \texttt{Exoformer}, obtaining a second set of posterior distributions.\par
    The posterior distributions obtained with our tool agree with those obtained from a Bayesian framework within $1\sigma$ for both planets (except for WASP-39b CH$_4$, which is compatible within $1.5\sigma$; see Fig.~\ref{fig-appendix:WASP-39b-taurex-vs-exoformer} and Fig.~\ref{fig-appendix:WASP-17b-taurex-vs-exoformer} in green and blue, respectively). This shows that \texttt{Exoformer}'s retrieval capability is comparable to TauREx's, although with slightly lower accuracy. However, we highlight an important trade-off between accuracy and computational speed: while TauREx yields precise results, its execution time is significantly longer than \texttt{Exoformer}'s, which accomplished the same task in a fraction of the time. TauREx required $\approx 498 $ h (WASP-39b) and $\approx 86 $ h (WASP-17b) with uniform priors to complete the retrievals on a single CPU core. By contrast, \texttt{Exoformer} inference to generate the priors itself took $\sim 2$ minutes on an NVIDIA A2 GPU.
    
    The retrieval performed on the two transmission spectra serves as a robustness test for \texttt{Exoformer}. Real observations often contain atmospheric phenomena unseen during the training phase. For example, WASP-39b's atmosphere contains strong traces of SO$_2$ and H$_2$S, which originate from photochemical processes \citep{Constantinou2023}. These unknown chemical species can interfere with the target molecules. Furthermore, clouds and haze can significantly affect retrievals by reducing or eliminating absorption features across observed wavelengths \citep{Lu2023}, ultimately resulting in biased measurements. Despite the challenges posed by real-world observations, the posterior distributions recovered by \texttt{Exoformer} remain consistent with those from TauREx, highlighting the robustness and reliability of our tool when applied to JWST spectroscopic data.

    \section{Informative priors}
    \label{section:informative-priors}

    The results obtained with \texttt{Exoformer}, as detailed in Section \ref{sec:comparisontaurex-exoformer}, present an opportunity to explore a hybrid approach that combines both the robustness and accuracy of Bayesian methods with the speed of deep learning, potentially enhancing the performance of existing Bayesian tools. Bayesian algorithms benefit from informative prior distributions, which accelerate convergence by constraining the probability within specific parameter space regions \citep{Gelman2017}.
    
    In fact, by reducing the volume of the prior distribution $V_{\text{prior}}$ over the volume of the posterior distribution $V_{\text{posterior}}$, the Kullback-Leibler (KL) divergence between the distributions is reduced \citep{Petrosyan2022}:
    \begin{equation}
        D_{KL}\big(\text{posterior}~||~\text{prior}\big)\approx\ln\biggl(\frac{V_{\text{prior}}}{V_{\text{posteriors}}}\biggl).
    \end{equation}
    Because the time complexity $T$ of the nested sampling algorithm is proportional to the KL divergence between priors and posteriors \citep{Petrosyan2022} then
    \begin{equation}
        T \propto \ln\biggl(\frac{V_{\text{prior}}}{V_{\text{posterior}}}\biggl).
    \end{equation}
    So when we restrict the priors space using informative priors, the overall effect is a reduction of the run-time of the algorithm. 

    \subsection{WASP-39b and WASP-17b}

\begin{figure*}[h]
    \sidecaption
      \centering
      \includegraphics[trim={0.2cm 0.2cm 0.5cm 0.5cm},clip,width=12cm]{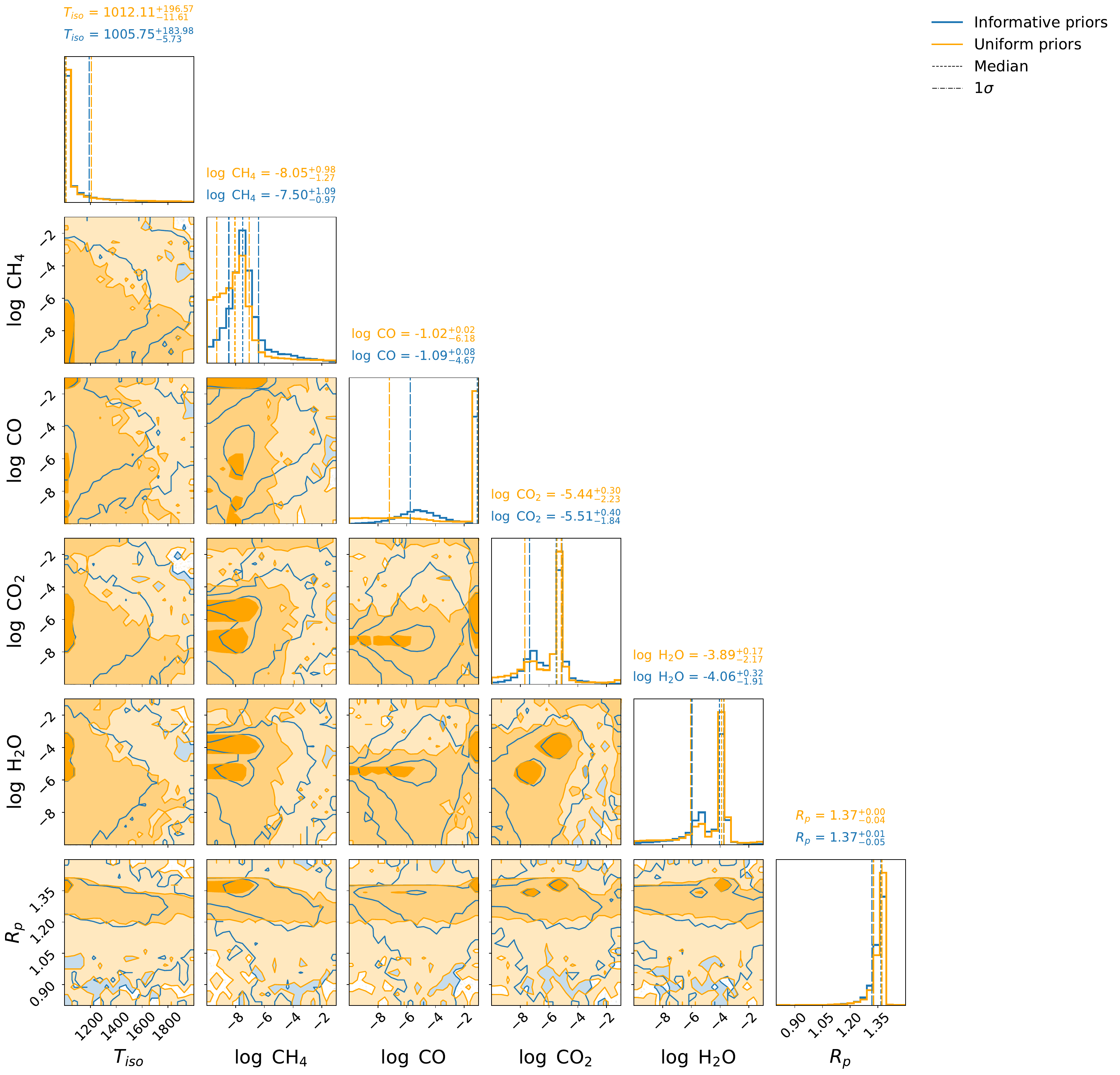}
      \caption{Corner plot for the WASP-39b retrieval. The posterior distributions obtained with informative priors are shown in blue, while those obtained with uniform priors are shown in orange. The dashed lines indicate the median of the distributions, while the dashed-dotted lines indicate the $1\sigma$ intervals. All the parameters from the two retrievals are compatible within $1\sigma$.}
      \label{fig:WASP-39b-retrieval}
    \end{figure*}

\begin{figure*}[h]
    \sidecaption
      \centering
      \includegraphics[trim={0.2cm 0.2cm 0.5cm 0.5cm},clip,width=12cm]{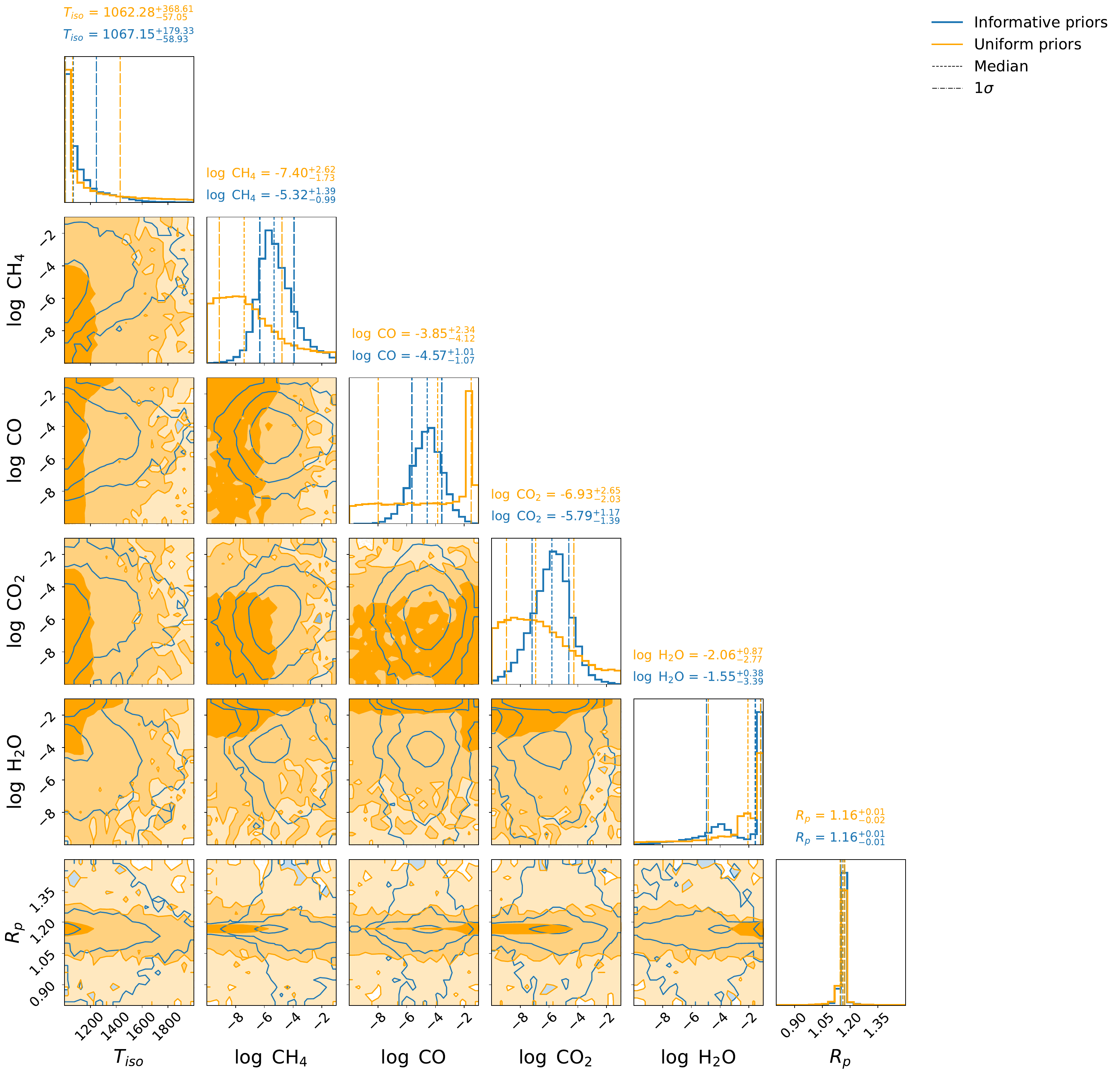}
      \caption{Corner plot for the WASP-17b retrievals, with labels as in Figure~\ref{fig:WASP-39b-retrieval}. The two retrievals are compatible with one another within $1\sigma$, with more regular posterior distributions from the informative priors.}
      \label{fig:WASP-17b-retrieval}
    \end{figure*}

   To assess the speedup a Bayesian retrieval could achieve with informative priors, we reexamined the WASP-39b and WASP-17b transmission spectra. We transformed \texttt{Exoformer}'s posterior distributions into informative priors for the \texttt{nestle} plugin, which can also be used with the \texttt{multinest} plugin \citep{Feroz_2009}. To keep the informative prior distributions consistent with the uniform priors used in the retrievals, we limited the informative priors to the same boundaries presented in Section \ref{sec:comparisontaurex-exoformer} for all parameters. The two retrieval methodologies returned a set of posterior distributions compatible with one another within $1\sigma$ (Fig.~\ref{fig:WASP-39b-retrieval} and Fig.~\ref{fig:WASP-17b-retrieval}). However, we observe significant improvement in the computational times (Table \ref{table:uniformvscustompriors}) for both planets: the speedup is close to eight times for WASP-39b and three times for WASP-17b. \par
\begin{figure*}[t]
    \sidecaption
      \centering
      \includegraphics[trim={0.4cm 0.1cm 0.0cm 0.2cm},clip,width=12cm]{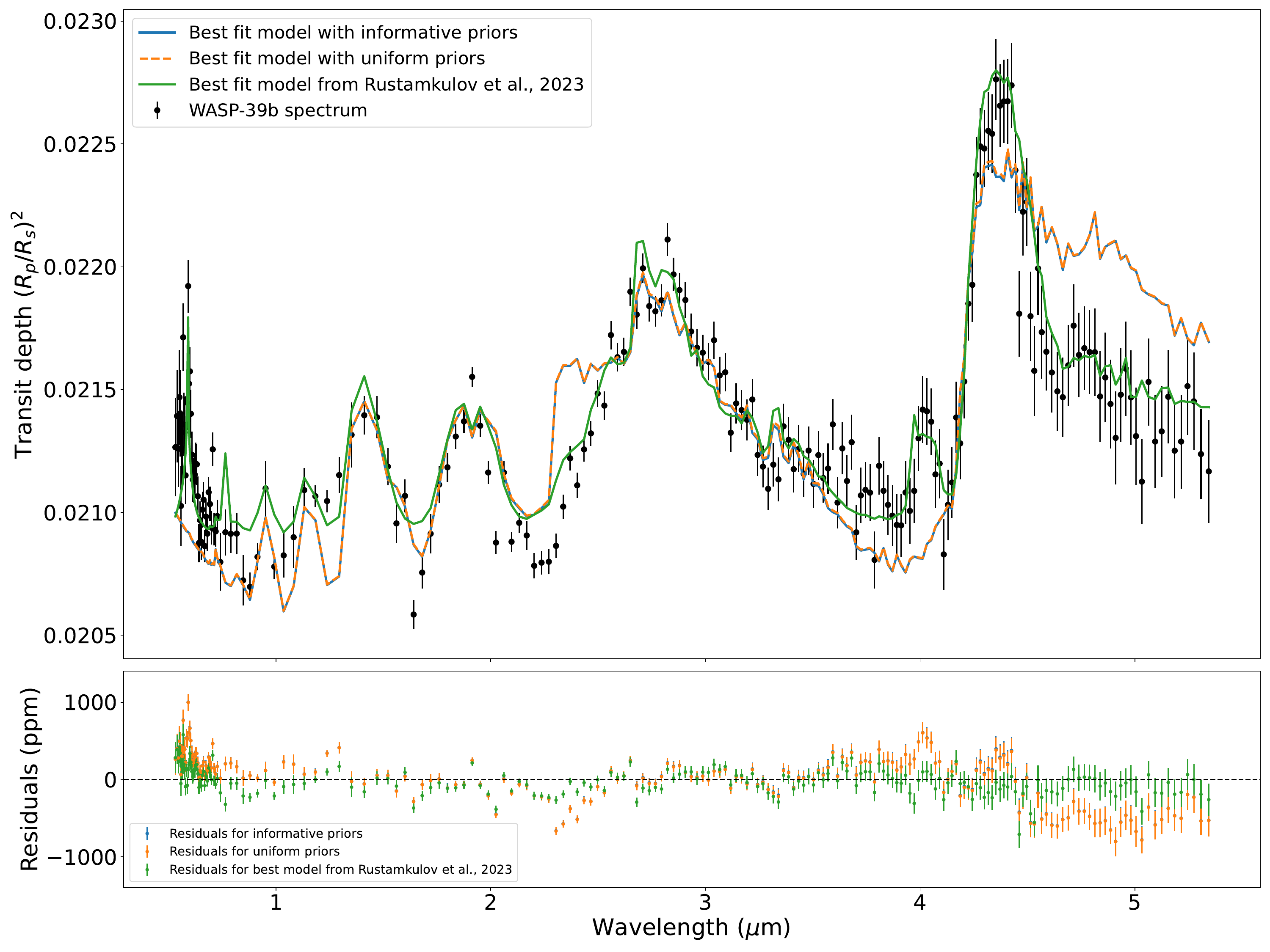}
      \caption{Best-fit models for the WASP-39b NIRSpec PRISM observation obtained using uniform (orange line) priors, informative (blue line) priors, and the 1D RCTE model (green line) from \citet{Rustamkulov2023}. Both the uniform and informative models miss important spectral features that are instead captured by the RCTE model.}
      \label{fig:wasp39b_best-fit}
    \end{figure*}
\begin{figure*}[t]
    \sidecaption
      \centering
      \includegraphics[trim={0.4cm 0.1cm 0.0cm 0.2cm},clip,width=12cm]{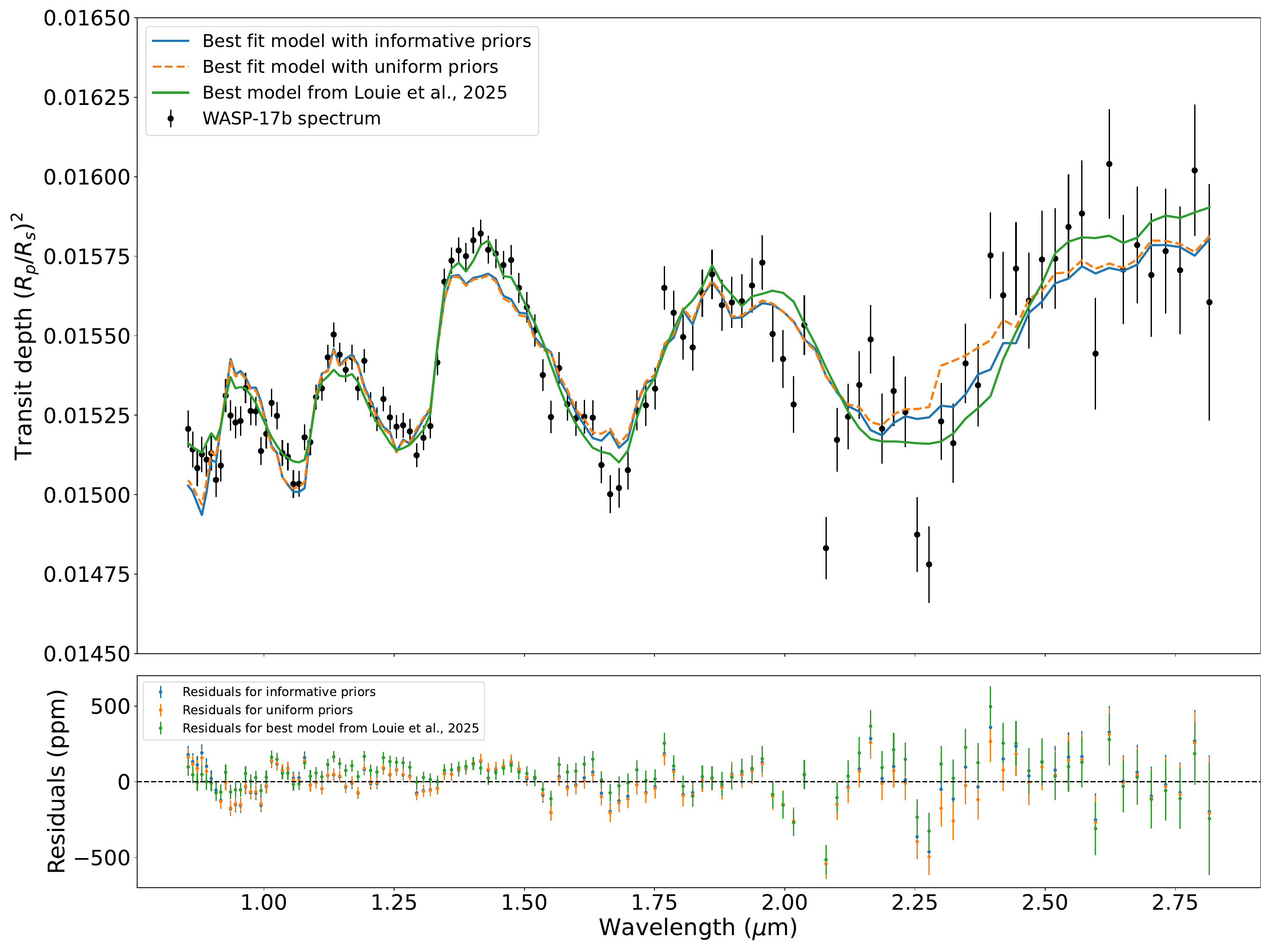}
      \caption{Best-fit models for the WASP-17b NIRISS SOSS observation obtained using uniform (orange line) and informative (blue line) priors compared to the best-fit model by \citet{Louie2025} (green line). The residuals show that our two models are consistent, differing only in the $2.25-2.5$ $\mu m$ wavelength range. As for WASP-39b, our atmospheric model cannot describe all spectra features, such as the strong H$_2$O features at $\sim 1.4, \sim 1.8$, and above $2.5$ $\mu m$.}
      \label{fig:wasp17b_best-fit}
    \end{figure*}

   The logarithmic Bayes factors \citep{kass_95, trotta_2007} for the two planets show different values (Table \ref{table:uniformvscustompriors}). These factors were computed as the difference between the evidence of the model with uniform priors and that with informative priors. For the WASP-39b retrieval, a log-Bayes factor $|\log{B}|=1.16 < 2$ \citep{kass_95, trotta_2007} indicates a weak preference for the model with informative priors. Indeed, the two retrievals show very similar best-fit models (Fig.~\ref{fig:wasp39b_best-fit}), with no evident differences. Furthermore, both models exhibit a clear deviation from observed data in the same wavelength range. This fitting yields higher transit depth values, explaining why the CO posterior distribution is pushed toward the upper boundary of the prior space. In the WASP-17b retrieval, a log-Bayes factor of $2 < |\log{B}|=4.87 < 5$ moderately favors \citep{kass_95, trotta_2007} the model found using uniform priors. Again, Fig.~\ref{fig:wasp17b_best-fit} shows very similar best-fit models, with an appreciable difference only in the range $2.25-2.5$ $\mu m$. This results in a greater abundance of H$_2$O for the uniform priors retrieval (Fig.~\ref{fig:WASP-17b-retrieval}). \par
   This tendency of the retrievals toward the model obtained with uniform priors -- despite the similar fit with the model using informative priors -- can be explained by the definition of Bayesian evidence itself. In fact, when two models both fit the data well, broader priors in one of them (such as those generated using uniform distributions) can yield larger evidence values \citep{Trotta_2008}, thereby increasing the absolute value of the Bayes factor. \par
   Figure~\ref{fig:WASP-39b-retrieval} (for CH$_4$) and Fig.~\ref{fig:WASP-17b-retrieval} lastly demonstrate the regularization effects of the Gaussian-like priors of \texttt{Exoformer}. As a result, the posterior distributions are smoother and more regular than those obtained with uniform priors \citep{Llorente_2023}.

   It is crucial to note that these two retrievals do not provide new solutions for the atmospheric characterization of the two selected exoplanets, but rather demonstrate a method that is consistent with -- and effectively enhances -- Bayesian retrievals. This is because the incompleteness of our atmospheric model can induce degenerate solutions. In the WASP-39b transmission spectrum, our atmospheric model (Section \ref{sec:dataset}) does not completely capture the prominent CO$_2$ feature at $\sim 4.3$ $\mu m$ \citep{Rustamkulov2023} (Fig.~\ref{fig:wasp39b_best-fit}) and shows a significant discrepancy with the observed data between $4.5$ and $5.5$ $\mu m$. Similarly, the strong H$_2$O features at $\sim1.4, \sim1.8$, and above $2.5$ $\mu m$ in WASP-17b \citep{Louie2025} (Fig.~\ref{fig:wasp17b_best-fit}) are not captured by our model. The introduction of a radiative-convective thermochemical equilibrium (RCTE) model (as in \citet{Rustamkulov2023} and \citet{Louie2025}) constrains the sampler to converge to more realistic solutions (green lines in Fig.~\ref{fig:wasp39b_best-fit} and Fig.~\ref{fig:wasp17b_best-fit}).\par
   Retraining \texttt{Exoformer} with a more sophisticated physical model is feasible but beyond the scope of this work, as it would require both implementing the new model in TauREx and generating a new, large-scale dataset. Our goal rather is to show that this hybrid approach is fully compatible and consistent with traditional retrievals using uniform priors, while also significantly reducing computational time.

   \subsection{Simulated planets}
    
   We also tested our strategy of combining \texttt{Exoformer} and TauREx with four additional simulated NIRSpec PRISM spectra, generated as described in Section \ref{subsec:testretrieval}. The last four rows of Table \ref{table:uniformvscustompriors} summarize the log-Bayes factors obtained from the retrievals on these simulated observations, indicating no strong evidence \citep{kass_95, trotta_2007} in favor of either retrieval method; thus, we can consider both implementations equivalent. However, in terms of computational timescales (Table \ref{table:uniformvscustompriors}), we notice a different behavior compared to the results from real observations: the retrievals with uniform priors are only slightly slower than those with informative priors. 
   
   The small improvement can be interpreted in the context of Bayesian inference. In our case, the four simulated transmission spectra were generated with the same forward model (Section \ref{sec:dataset}) used in the retrieval process, with realistic instrumental noise added through \texttt{Pandexo}. Thus, we describe the simulated observations with exactly the number of parameters necessary to explain the physical processes behind their formation. Following \citet{Trotta_2008}, our parameters are already well constrained by the data (the effective number of parameters, given by the Bayesian complexity \citep{Spiegelhalter_2002} equals the number of free parameters), so any additional information provided by the informative priors contributes only marginally to the model's predictivity. As a result, informed prior knowledge yields a minor speedup compared to using uniform priors.

   In contrast, JWST observations of real exoplanets require additional parameters to explain the additional spectral features (see Fig.~\ref{fig:wasp39b_best-fit} and Fig.~\ref{fig:WASP-17b-retrieval}), which are not provided by the model used in our retrieval. As a consequence, in this case, the retrieval performance is controlled by the parameter space available from the priors, and so how much the Bayesian algorithm has to explore.

   \begin{table*}[t]
       \def\arraystretch{1.5}
       \centering
       \caption{Summary table for the six atmospheric retrievals performed in this work (two on real JWST data and four on simulated observations).}
       \begin{tabular}{l c c c c c c}
       \hline\hline
       {Planet} & \multicolumn{2}{c}{Timescales} & {Speedup} & \multicolumn{2}{c}{{log-Bayesian Evidence}} & {log-Bayes factor} \\
       \hline
        & {Uniform Prior} & {Informative Prior} & & {Uniform Prior} & {Informative Prior} &\\
       \hline
        WASP-39b & 498 h & 64 h & 7.8 & 668.04$\pm$0.14 & 669.20$\pm$0.14 & -1.16 \\
        WASP-17b & 86 h & 27 h & 3.2 & 413.52$\pm$0.10 & 408.64$\pm$0.11 & 4.87 \\
        Simulated 1 & 44 h & 36 h & 1.2 & 2185.67$\pm$0.11 & 2185.21$\pm$0.11 & 0.46 \\
        Simulated 2 & 71 h & 65 h & 1.1 & 2175.64$\pm$0.11 & 2174.42$\pm$0.11 & 1.22 \\
        Simulated 3 & 33 h & 27 h & 1.2 & 2189.13$\pm$0.10 & 2191.779$\pm$0.09 & -2.649 \\
        Simulated 4 & 23 h & 20 h & 1.1 & 2183.214$\pm$0.08 & 2181.244$\pm$0.09 & 1.97 \\
       \hline
       \end{tabular}
       \tablefoot{The first three columns show the computational timescales for the retrievals with uniform and informative priors, along with the corresponding speedup. The last two columns show the log-Bayesian evidences for the same retrievals and the corresponding log-Bayes factor (computed following \citet{kass_95, trotta_2007} as the difference between evidences with uniform and informative priors).}
       \label{table:uniformvscustompriors}
   \end{table*}

\section{Discussion and conclusions}

    Atmospheric retrievals frequently rely on uninformative priors, such as uniform distributions, due to limited knowledge of molecular abundances and temperatures in exoplanetary atmospheres. This approach forces Bayesian frameworks, such as nested sampling, to explore the entire parameter space, creating a computational bottleneck that becomes increasingly predominant as the number of fitting parameters increases. However, when prior knowledge is available, we can avoid excessive sampling in regions of the parameter space with improbable values \citep{Ashton_2022}. Deep learning models such as \texttt{Exoformer} address these challenges by extracting useful information from the training data and constraining the range of possible values for the fitting parameters. These tools rapidly compute parameter distributions — typically in minutes — that a Bayesian framework can use to focus on high-probability regions of the parameter space, resulting in a more efficient and smoother inference than with uniform priors \citep{Gelman2017}.

    A key drawback of this strategy is the unrepresentative prior problem \citep{Chen_2019,Chen2023} that affects nested sampling-based algorithms. As the prior distribution moves away from the true value of the parameter, the likelihood remains almost flat within the prior space \citep{Chen_2019,Chen2023}. By definition, the nested sampling algorithm selects live points from priors with higher likelihood values at each iteration \citep{Skilling2006}. Consequently, on a flat likelihood surface, the time spent searching for higher-likelihood values is significantly greater. This results in slow convergence timescales or -- in the worst case -- trapping of the algorithm in low-likelihood regions, yielding incorrect posterior distributions \citep{Chen_2019,Chen2023}. Our tool, however, demonstrates robustness by deriving the correct fitting parameters (within $1\sigma$), thereby minimizing the risk of exploring incorrect regions of the prior space.
    
    Our strategy of computing informative priors through a transformer-based tool and combining it with a Bayesian retrieval proved effective in reducing the computational timescales of atmospheric retrievals. We achieved a three to eight time speedup compared to classical retrievals performed using uniform prior distributions, while maintaining consistency with the retrieved parameters and the best-fit models.

    The transformer architecture at the foundation of our tool has been effective in regression and classification tasks for sequential data. However, further improvements to the dataset are necessary to prepare \texttt{Exoformer} for large future atmospheric surveys such as Ariel, which will give us deeper insights into atmospheres for a wide population of exoplanets \citep{Edwards_2019}. For WASP-39b and WASP-17b, our model fails to fit all spectral features, indicating the absence of molecules and physical processes observed by JWST. When describing populations of hot Jupiters, these failures can be address by including additional molecular compounds typically found in hot Jupiters (sulfur compounds such as SO$_2$ and H$_2$S \citep{Zahnle_2009}, or oxidized compounds such as TiO, VO, and SiO \citep{Desert_2008}), multidimensional effects (e.g., \citet{Helling_2020, Zingales_Falco_Pluriel_Leconte_2022}), or different chemical states (e.g., chemical disequilibrium and photochemistry \citep{Tsai_2021}).
    
    Another aspect to consider is the type of exoplanets described by the dataset: the hot giant planets in our dataset constitute only a small fraction of the broader exoplanet family. It has been well established that small exoplanets are the most abundant class \citep{Petigura2013}. However, their atmospheres occupy different regions of the parameter space (smaller masses and radii, higher bulk densities, and different physical processes; e.g., \citet{Madhusudhan_2016}) than those of giant exoplanets. \par
    A comprehensive deep learning tool for future survey analysis should also be able to analyze the atmospheric spectra of these small planets. The advantage of our data-driven approach is that the architecture remains fixed while training \texttt{Exoformer} on new datasets. The updated network state can then be saved and applied to the appropriate data.

    The custom wavelength grid, derived from \citet{Zingales2018}, represents another critical point. Although the grid covers the wavelength domain of JWST filters, its resolution was initially designed for HST observations. Consequently, it is no longer adequate for application with high-resolution JWST data and for future observations with Ariel. In fact, during the interpolation, some spectral information could be lost because of the lower number of points in the grid compared to the observations. A new custom grid should be developed with a sufficiently fine set of wavelength points to match the typical resolution of JWST instruments. In this way, we will help \texttt{Exoformer} extract more spectral features from real JWST observations, further increasing its accuracy.

\section*{Data availability}
    The \texttt{Exoformer} and \texttt{ExoformerPriors} TauREx plugin are available on GitHub respectively at \href{https://github.com/antares07/Exoformer.git}{Exoformer} and \href{https://github.com/antares07/ExoformerPriorsTaurex.git}{ExoformerPriorsTaurex}.

\begin{acknowledgements}
      This publication was produced while attending the PhD program in Astronomy at the University of Padova, Cycle XXXIV, with the support of a scholarship co-financed by the Ministerial Decree no. 118 of 2nd March 2023,760 based on the NRRP - funded by the European Union - NextGenerationEU - Mission 4 Component 1 – CUP C96E23000340001. GPi and GMa acknowledge support by the Space It Up project funded by the Italian Space Agency, ASI, and the Ministry of University and Research, MUR, under contract n. 2024-5-E.0 - CUP n. I53D24000060005. TZi acknowledges support from CHEOPS ASI-INAF agreement n. 2019-29-HH.0, NVIDIA Academic Hardware Grant Program for the use of the Titan V GPU card and the Italian MUR Departments of Excellence grant 2023-2027 “Quantum Frontiers”.
\end{acknowledgements}

\bibliographystyle{aa}
\bibliography{bibliography}

\begin{appendix}
\clearpage

\clearpage
\onecolumn

\section{Additional figures}
    \begin{figure*}[!htbp]
      \centering
      \includegraphics[width=\textwidth]{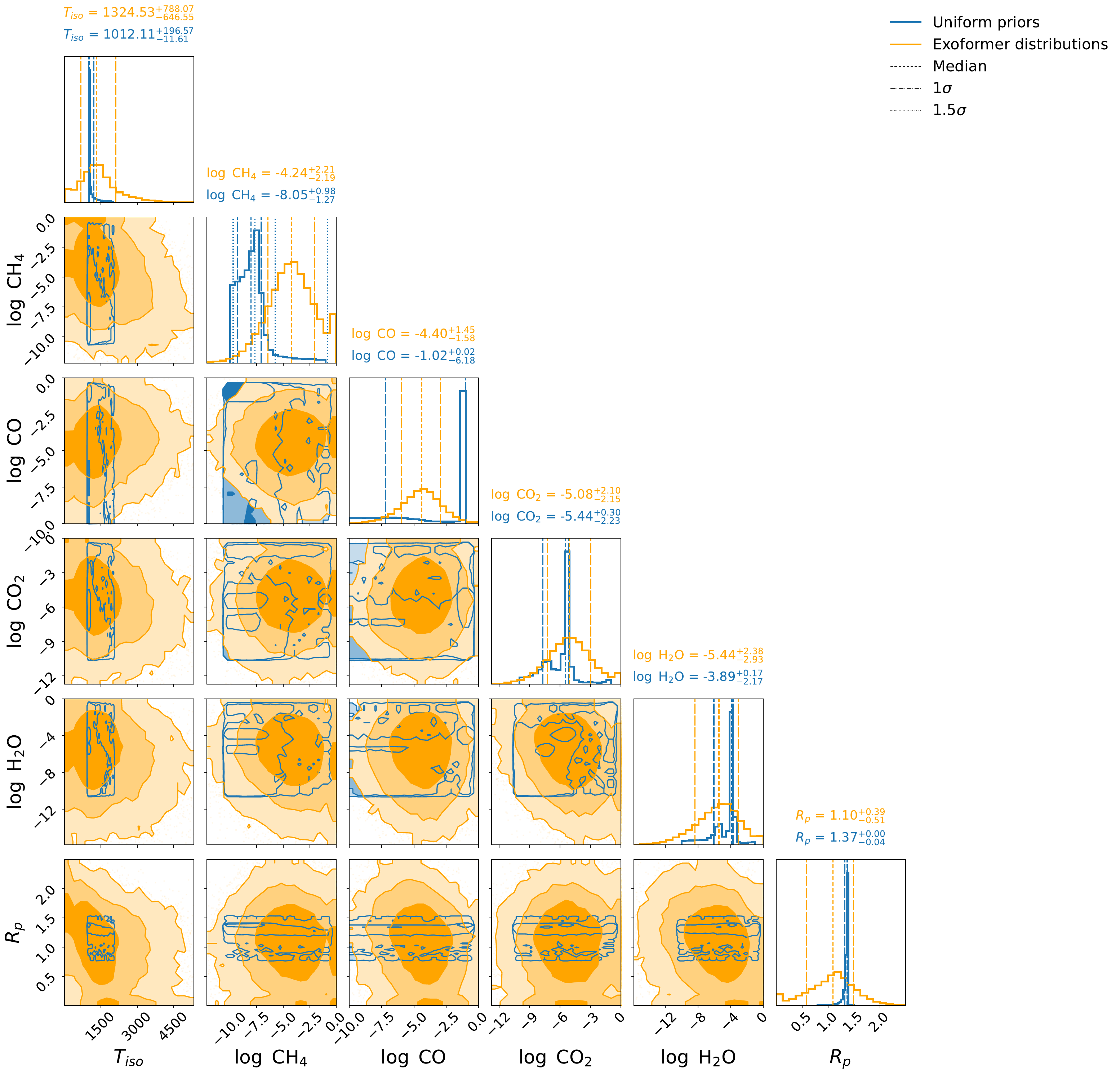}
      \caption{Corner plot for the WASP-39b retrieval. In blue we show the posterior distributions retrieved with TauREx using uniform priors, while in orange we show the distribution obtained using \texttt{Exoformer}. The distributions are compatible with one another within $1\sigma$, while for CH$_4$ within $1.5\sigma$.}
      \label{fig-appendix:WASP-39b-taurex-vs-exoformer}
    \end{figure*}
    \begin{figure*}[!htbp]
      \centering
      \includegraphics[width=\textwidth]{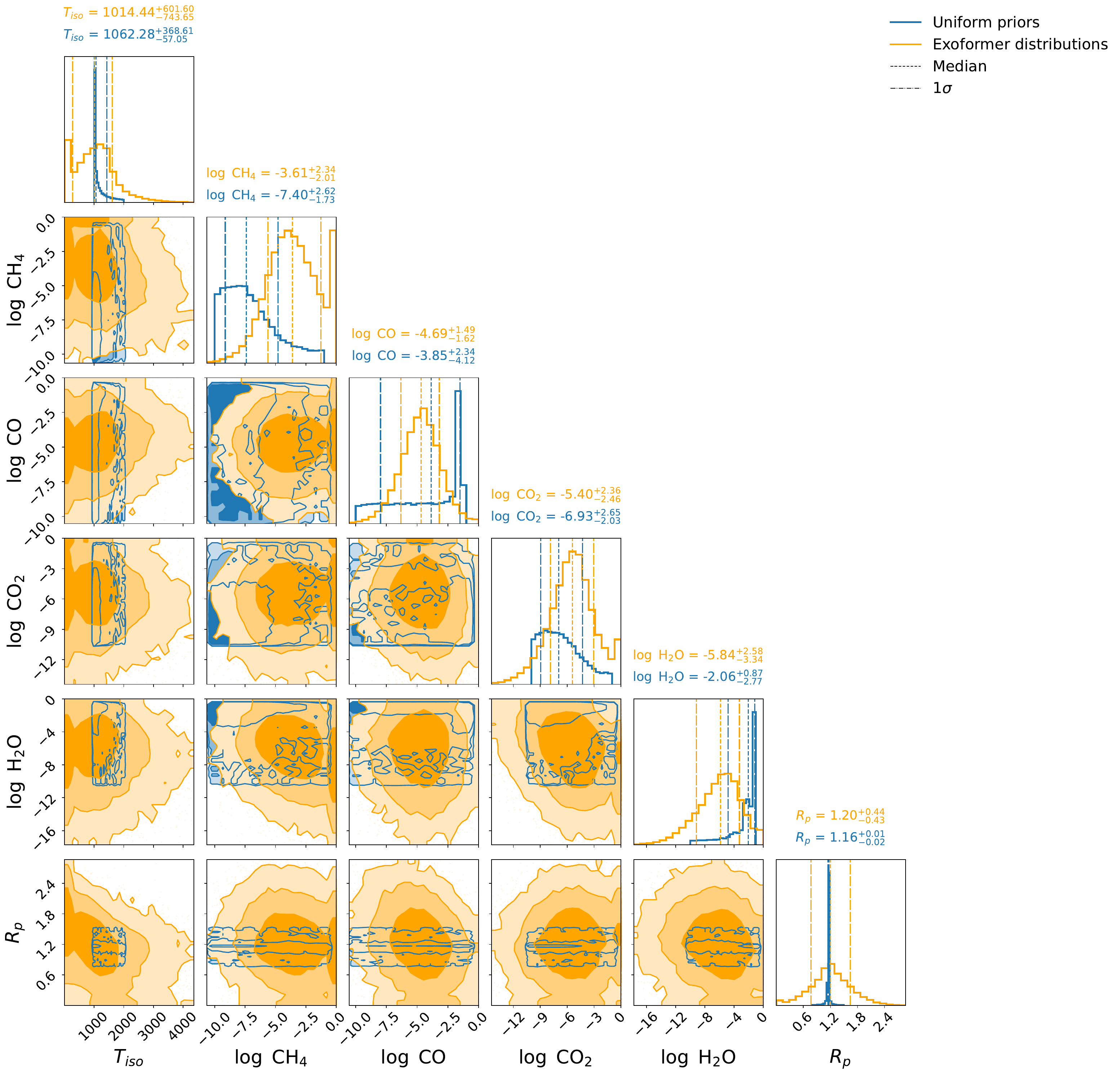}
      \caption{Corner plot for the WASP-17b retrieval. In blue we show the posterior distributions retrieved with TauREx using uniform priors, while in orange we show the distribution obtained using \texttt{Exoformer}. The distributions are compatible with one another within $1\sigma$.}
      \label{fig-appendix:WASP-17b-taurex-vs-exoformer}
    \end{figure*}
\end{appendix}

\end{document}